\documentclass[twocolumn]{aastex62}
\usepackage[utf8]{inputenc}

\submitjournal{ApJ}

\shorttitle{The Prospects of Observing TDEs with the LSST}
\shortauthors{Bricman and Gomboc}

\begin{document}

\title{The Prospects of Observing Tidal Disruption Events with the LSST}

\correspondingauthor{Katja Bricman}
\email{katja.bricman@ung.si}

\author{Katja Bricman}
\affil{Center for Astrophysics and Cosmology \\
University of Nova Gorica\\
Vipavska 13 \\
SI-5000 Nova Gorica, Slovenia}

\author{Andreja Gomboc}
\affil{Center for Astrophysics and Cosmology \\
University of Nova Gorica\\ 
Vipavska 13 \\
SI-5000 Nova Gorica, Slovenia}

\begin{abstract}

The upcoming Large Synoptic Survey Telescope (LSST) will observe 18 000 deg$^2$ of the Southern sky and is expected to discover thousands of transients every night due to its large coverage area and its observing strategy. In this work we address the prospects for the LSST in discovering Tidal Disruption Events (TDEs) and in probing the supermassive black hole (SMBH) mass distribution in the Universe. We used the LSST simulation framework and defined TDE catalogs on 20 fields of 20.25 deg$^2$ size. TDE properties were defined by randomly chosen impact factors and SMBH masses drawn from six different mass distributions. Observations of TDEs over 10 years of LSST operation were simulated by querying the simulated observing strategy database \texttt{minion\_1016}. Based on the results of our simulations we estimate that the LSST should discover between 35 000 and 80 000 TDEs in 10 years of operation, depending on the assumed SMBH mass distribution. We also find that probing the SMBH mass distribution with TDE observations will not be straightforward due to the fact that TDEs caused by low mass black holes ($10^ 5 M_\odot$) are expected to be less luminous and shorter than TDEs by heavier SMBHs ($> 10^6 M_\odot$), and therefore will mostly be missed by the irregular LSST cadence \texttt{minion\_1016}.
\end{abstract}


\section{Introduction}
\label{introduction}

When a star in the nucleus of a galaxy gets scattered into an unfortunate orbit leading it close to the supermassive black hole (SMBH) in the center of its host, the star can be torn apart by the black hole's strong tidal forces \citep{Rees:1988bf, Phinney1989, Evans:1989qe}. This process, known as a Tidal Disruption Event (TDE), emits a bright flare of light, which then decays on time scales from months to years.

The majority of SMBHs found in centers of galaxies are quiescent and therefore generally very hard to study. However, TDEs are recognized as one of the most promising phenomena in the study of non-active SMBHs. The observed emission depends on different parameters concerning the objects and orbital dynamics involved, such as; the mass of the black hole, the mass, radius, and structure of the star, and the distance from the black hole at which the star gets disrupted \citep{Kochanek:1993cm, Gomboc:2005wu, Lodato:2008fr, Strubbe:2009qs, Lodato:2010xs, Guillochon:2012uc, Mockler:2018xne}. Therefore, the observed light curves of such events can, at least in principle, provide us with information about the disrupted stars, as well as SMBHs responsible for the events.

TDEs are very rare, with only around 70 candidates discovered so far \citep[e.g.][]{vanVelzen:2010jp, Gezari:2012sa, Arcavi:2014iha,  Chornock:2013jta, Holoien:2014jha, Holoien:2015pza, Holoien:2016uaf, Leloudas:2016rmh, Wyrzykowski:2016acu, Blagorodnova:2017gzv, Gezari:2017qnq, Holoien:2018huf, Holoien:2018oby, vanVelzen:2018dwv, Leloudas:2019tfg, Holoien:2019zry}, however this number has been continuously increasing over the past few years due to discoveries by time-domain surveys, such as Palomar Transient Factory (PTF; \citealt{Law:2009ys}), All-Sky Automated Survey for Supernovae (ASAS-SN; \citealt{Shappee:2013mna, 2017PASP..129j4502K}), the Asteroid Terrestrial-impact Last Alert System (ATLAS; \citealt{2018PASP..130f4505T}), and the Zwicky Transient Facility (ZTF; \citealt{2019PASP..131f8003B}). Currently, TDEs are being discovered at a rate of approximately 10 per year. 

The rate at which stars in the cores of galaxies are disrupted depends on their density and scattering mechanisms. Dynamical models of stellar orbits in central regions of galaxies predict that the rate of TDEs is $10^{-4}$ per galaxy per year \citep{Magorrian:1999vm, Wang:2003ny}, while the observed sample implies that the rate of TDEs is $10^{-5}$ per galaxy per year \citep[e.g.][]{vanVelzen:2014dna, Holoien:2015pza}. Due to this low rate, large surveys monitoring hundreds of thousands of galaxies, such as the future Large Synoptic Survey Telescope (LSST), will be crucial in enlarging the observed TDE sample size. 

The LSST \citep{Ivezic:2008fe} is an upcoming sky survey project, which will conduct a 10-year long survey of the dynamic Universe in six optical bands, \emph{u, g, r, i, z} and \emph{y}, covering the wavelength range between $320$ and $1050$ nm. With its large field of view of $9.6$ deg$^2$ it will be able to cover around 10 000 deg$^2$ of sky each night, and therefore be capable of mapping the entire visible sky in just a few nights. The primary mirror will measure $8.4$ m in diameter, which will allow imaging to very faint magnitudes, up to 24.4 in \emph{r} band in a single exposure. The combination of all this will result in the mapping of tens of billions of stars and galaxies, and by doing so, create a multi-color view of the Universe \citep{Abell:2009aa, Ivezic2013, Gressler2016}.

According to the cadence proposed in \cite{Ivezic2013}, the survey will continuously monitor 18 000 deg$^2$ of the visible sky in the Southern hemisphere, and each field will be visited around 900 times over the 10 years survey duration. This will enable studies of small objects in the Solar System, the structure of the Milky Way, galactic evolution, variable and transient sources, properties of dark matter and dark energy, and discoveries of yet unknown astrophysical objects. Images obtained with the LSST will be analyzed in real-time in order to identify objects which might have changed their brightness since the previous observation, or which might have moved. Therefore, the LSST will be a powerful tool in the search for transients, including TDEs.

In order to estimate the number of TDEs we may expect the LSST to detect, the quality of their light curve coverage, and whether it will be possible to use them to probe the SMBH mass distribution, we performed simulations using the LSST simulation framework \citep{Ivezic:2008fe, Connolly2010, Connolly2014, Delgado2016}. The framework includes all the components which may significantly affect observational data, from the design of the telescope, to conditions at the observing site, to the survey strategy. Since the simulation framework does not include TDEs, we imported them as a new type of objects. Our basic steps were the following: first, we randomly chose host galaxies and attributed them a central SMBH with a mass drawn randomly from an assumed SMBH mass distribution. Since the real SMBH mass distribution is still uncertain, particularly at the low-mass end, we considered 6 different distributions (assuming no evolution with redshift) in order to test their effect on the number of detected TDEs. The optical properties of each particular TDE depend on the mass of the SMBH and on the properties (mass, radius) of the star being disrupted, as well as on the penetration factor. We considered all stars to be Solar-like and assumed that TDEs occur at random times. We then calculated the spectral energy distributions (SEDs) of TDEs at different times after the disruption using \texttt{MOSFiT}, a model based on hydrodynamical simulations of TDE fallback rate \citep{Guillochon:2017bmg, Mockler:2018xne}. We imported these SEDs in the LSST simulation framework and reproduced LSST observations of TDEs over 10 years on 20 fields on the sky, each covering an area of $20.25$ deg$^2$.

This paper is organized as follows: in Section \ref{tdes} we describe briefly the theoretical background of TDEs, in Section \ref{smbhdistribs} we present the SMBH mass distributions used in our simulations, in Section \ref{simulations} we describe the simulation setup, and we present our results in Section \ref{results}. We give our conclusions in Section \ref{conclusions}.

\section{TDEs}
\label{tdes}

SMBHs with masses ranging from $10^5$ to $10^{10}$ Solar masses, are common in the nuclei of galaxies, including our own \citep{Phinney1989}. Since they do not emit light, they are generally very hard to study. By producing luminous flares in the cores of galaxies, TDEs pose an opportunity to detected dormant black holes residing in galactic centers.

\subsection{Dynamics}
\label{TDEdynamics}
Consider a star of mass $M_*$ and radius $R_*$ moving on a highly eccentric orbit around a SMBH of mass $M_{BH}$. If the distance of the closest approach to the black hole, the pericenter distance $r_{p}$, lies within the tidal sphere with radius
\begin{equation}
r_{t} = R_* \bigg( \frac{M_{BH}}{M_*} \bigg)^{1/3},
\label{rTidal}
\end{equation}
then the tidal forces of the black hole overcome the star's self gravity, and the star is ripped apart \citep{Rees:1988bf}. The penetration factor is defined as the ratio of two distances 
\begin{equation}
\beta =r_{t}/r_{p}.
\label{beta}
\end{equation}

To simplify the model, we assume the star is on a parabolic orbit. At distances $r \gg r_{t}$ the star can be approximated as a point source in the gravitational field of the black hole, however, when it approaches the tidal radius, its size becomes important. Different distances to the black hole, at which different fluid elements of the disrupting star lie, cause a sizable spread in specific orbital energy $\epsilon$ within the star. The parts furthest from the black hole have a positive specific binding energy, while the energy of parts closest to the black hole is negative \citep{Rees:1988bf}.

After the disruption, part of the stellar debris remains bound to the black hole, while other parts of the disrupted star escape its gravitational pull. The returning bound material feeds the black hole and produces a flare of light. In early models it was assumed that the bound material quickly circularizes and forms an accretion disk \citep{Rees:1988bf, Phinney1989}. The rate at which material returns and feeds the accretion disk is the fallback rate, $\dot{M}_{fb}$, and it follows, in this simple model, a $t^{-5/3}$ decline. The emitted luminosity is expected to follow this decline as well.

The multi-wavelength emission of the TDEs was analytically modeled based on the dynamics and the fallback rate \citep[see e.g.][]{Strubbe:2009qs, Lodato:2010xs}. Optical light curves in these models have two contributions: the outflow emission, which dominates in the early stages, and the emission of the accretion disk, which takes over at later stages.

While observed optical TDEs quite often follow a power-law decay, which seems to be consistent with $t^{-5/3}$ at early times, as shown in \cite{vanVelzen:2018dlv}, some recent observations challenge this simple picture. For example, observed events, such as ASASSN-14ae \citep{Holoien:2014jha}, ASASSN-14li \citep{Holoien:2015pza} or iPTF16fnl \citep{Blagorodnova:2017gzv}, all show discrepancies from the $t^{-5/3}$ decay. Furthermore, the observed TDEs are about one to two orders of magnitude brighter than analytical models of the optical light curves would suggest \citep{Gezari:2009en, vanVelzen:2010jp, Wevers:2017llv}. Contrary to this simple model, some TDEs show unusual light curve shapes, for example the UV re-brightening/flattening phases in the cases of TDE candidates ASASSN-15lh \citep{Leloudas:2016rmh}, PSK18h \citep{Holoien:2018oby} or AT 2018fyk \citep{Wevers:2019dxz}. In addition, some characteristics of TDEs, such as low black body temperatures, minimal color evolution, and high peak luminosities \citep{vanVelzen:2010jp, Arcavi:2014iha, Hung:2017lxm} are all inconsistent with the simple picture.

In recent years it has been argued; that the fallback rate depends on the structure of the star, on the accretion process of the stream of stellar debris into the black hole, and even on the black hole spin, all of which can cause a discrepancy from the $t^{-5/3}$ time evolution \citep{Lodato:2008fr, Guillochon:2012uc}. Furthermore, it was suggested that the circularization might be a much slower process than previously thought \citep{Shiokawa:2015iia, Dai:2015eua, Piran:2015gha, Guillochon:2015qfa, Hayasaki:2015pxa}.

Alternatively, numerical simulations have provided new results for the optical emission of TDEs. It has been proposed that stream-stream collisions produce shocks, which can dissipate the orbital energy and produce the observed emission \citep{Piran:2015gha, Krolik:2016grf, Bonnerot:2016cob}. It has also been suggested that the optical emission originates from the reprocessed X-ray emission in the stellar debris or disk or in an outflow \citep{Loeb:1997dv, Guillochon:2013jda, Miller:2015jda, Metzger:2015sea}. It remains unclear how exactly the fallback rate translates to the observed luminosity.

\subsection{MOSFiT SED model}
\label{mosfitSED}

We used \texttt{MOSFiT} \citep{Guillochon:2017bmg, Mockler:2018xne} to calculate the SEDs of TDEs at different times after the disruption. \texttt{MOSFiT} uses \texttt{FLASH} simulations of the fallback rate, and seems to describe previous observations of TDEs in optical wavelengths well. As shown in \cite{Mockler:2018xne}, fitting this model to observations enables the determination of some TDE parameters, such as, the mass of the black hole, the penetration factor, the stellar mass, the type of the disrupted star, and the peak time.

\begin{figure}
\includegraphics[width=\linewidth]{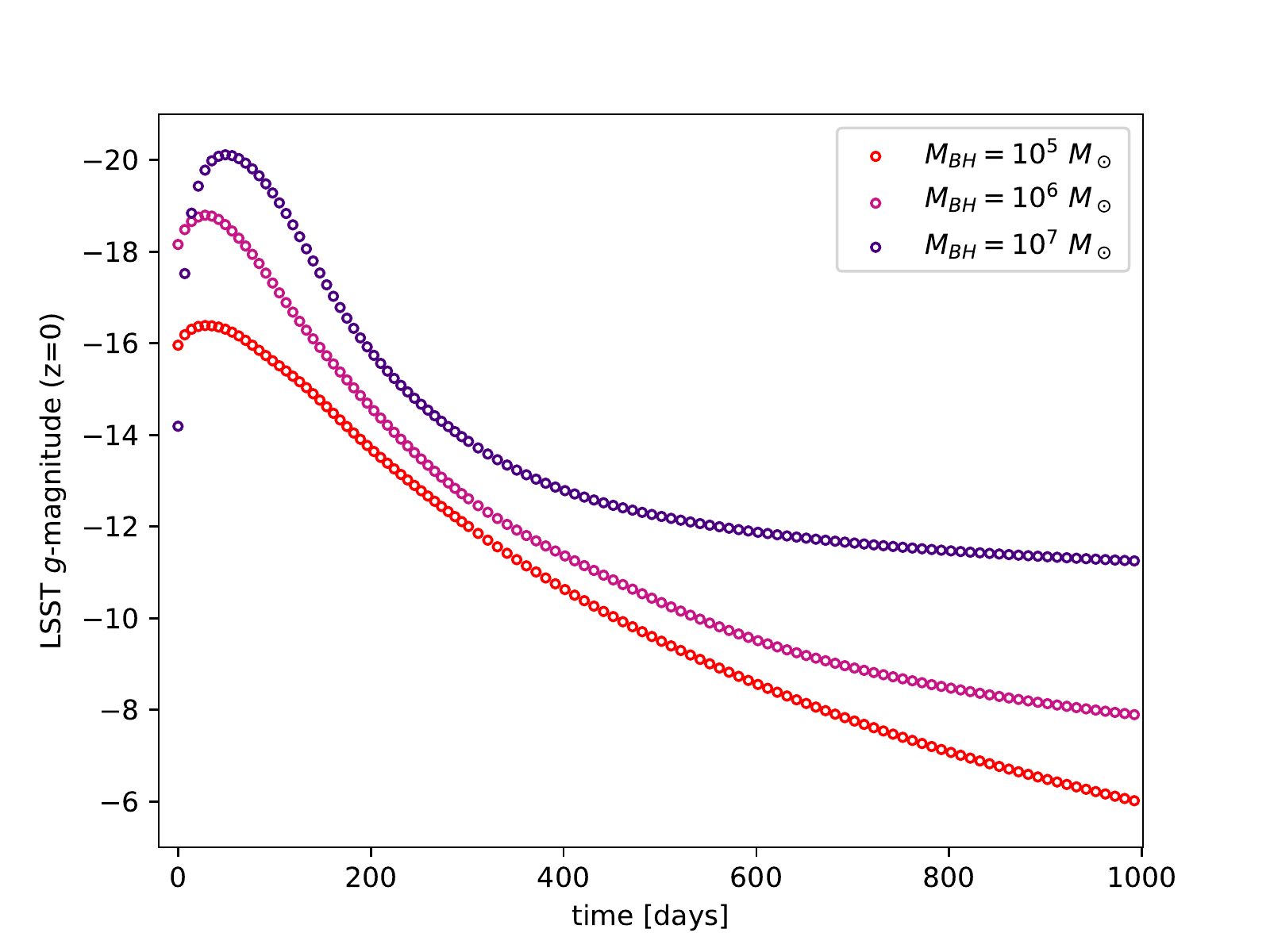} 
\caption{\texttt{MOSFiT} generated light curves of three TDEs with black hole masses: $10^5 \, M_{\odot}$ (red), $10^6 \, M_{\odot}$ (pink), and $10^7 \, M_{\odot}$ (violet). In all three events a Solar-type star disruption with $\beta=1$ was assumed. The absolute magnitudes were calculated in LSST \emph{g} band.} 
\label{fig:bh_comparison} 
\end{figure}

\texttt{MOSFiT}'s main purpose is to provide a tool for fitting transients. However, it can also be used to generate light curves and SEDs at any time after the disruption of a TDE with chosen parameters. Light curves of three events calculated in the LSST \emph{g} band using \texttt{MOSFiT} are shown in Figure \ref{fig:bh_comparison}.

Using \texttt{MOSFiT} we have created a library of SEDs for different events, where we varied two parameters: the black hole mass, and the penetration factor $\beta$. In all cases we used a one Solar mass star, described by a hybrid model which blends between polytropic models with $\gamma=5/3$ and $\gamma=4/3$ (see \citealt{Mockler:2018xne}). We placed all events at redshift $z=0$, in order to obtain the rest frame SEDs at a given time after the disruption. All other parameters from Table 1 in \cite{Mockler:2018xne} were kept constant.

Although, as shown in \cite{Kochanek:2016zzg}, TDEs by black holes of $\le 10^{7.5} \, M_\odot$ are dominated by stars of $M_* \sim 0.3 \, M_\odot$, the authors also argue that the mass function is relatively flat for $M_* \lesssim M_\odot$. Since we do not know exactly what the mass function for stars in galactic centers is, we chose not to vary the stellar mass and the polytropic index and to instead take a Solar-type star as a representative disruption candidate. This may affect our results, since less massive stars with different $\gamma$ will result in a different emission signature than the disruption of a $1 M_\odot$ star. Typically, less massive stars produce shorter and less luminous events. In particular we find that for less massive black holes ($M_{BH} < 10^{6} M_\odot$) the rest-frame peak luminosity of a 0.3 $M_\odot$ star is of approximately the same order as for a $1 M_\odot$ star. However, for a black hole of $10^7 M_\odot$ an event with a $0.3 M_\odot$ star is $\sim 3$ times less luminous than an event with a $1 M_\odot$ star. \cite{Mockler:2018xne} find that fitting a specific TDE with three different stars (with 0.1 $M_\odot$, 1 $M_\odot$ and 10 $M_\odot$) produces comparable results, and therefore the mass of the star does not have a significant effect on the black hole mass as determined by the fit. Thus, the SMBH mass distribution drawn from our simulated TDEs should not be significantly affected by changing the mass of the star in our simulations. Since brighter events tend to be easier to detect, taking only Solar-type stars in our simulations means that the number of detected TDEs we obtain should be considered as an upper limit.

We realize that \texttt{MOSFiT} can reproduce light curves of TDEs which are considered normal, but can have difficulties in reproducing the light curves of events with some unusual properties, such as UV re-brightening/flattening, as in the cases of ASASSN-15lh, PS18kh, or AT 2018fyk. However, re-brightening in UV should neither affect the optical light curves considerably nor the estimation of the SMBH mass. Since our goal is not to model individual TDEs light curves, but to obtain an estimation for the number of TDEs detected by the LSST, we use the simple light curve model without any unusual behaviours.

\section{SMBH distributions}
\label{smbhdistribs}

\begin{figure}
\centering
\includegraphics[width=\linewidth]{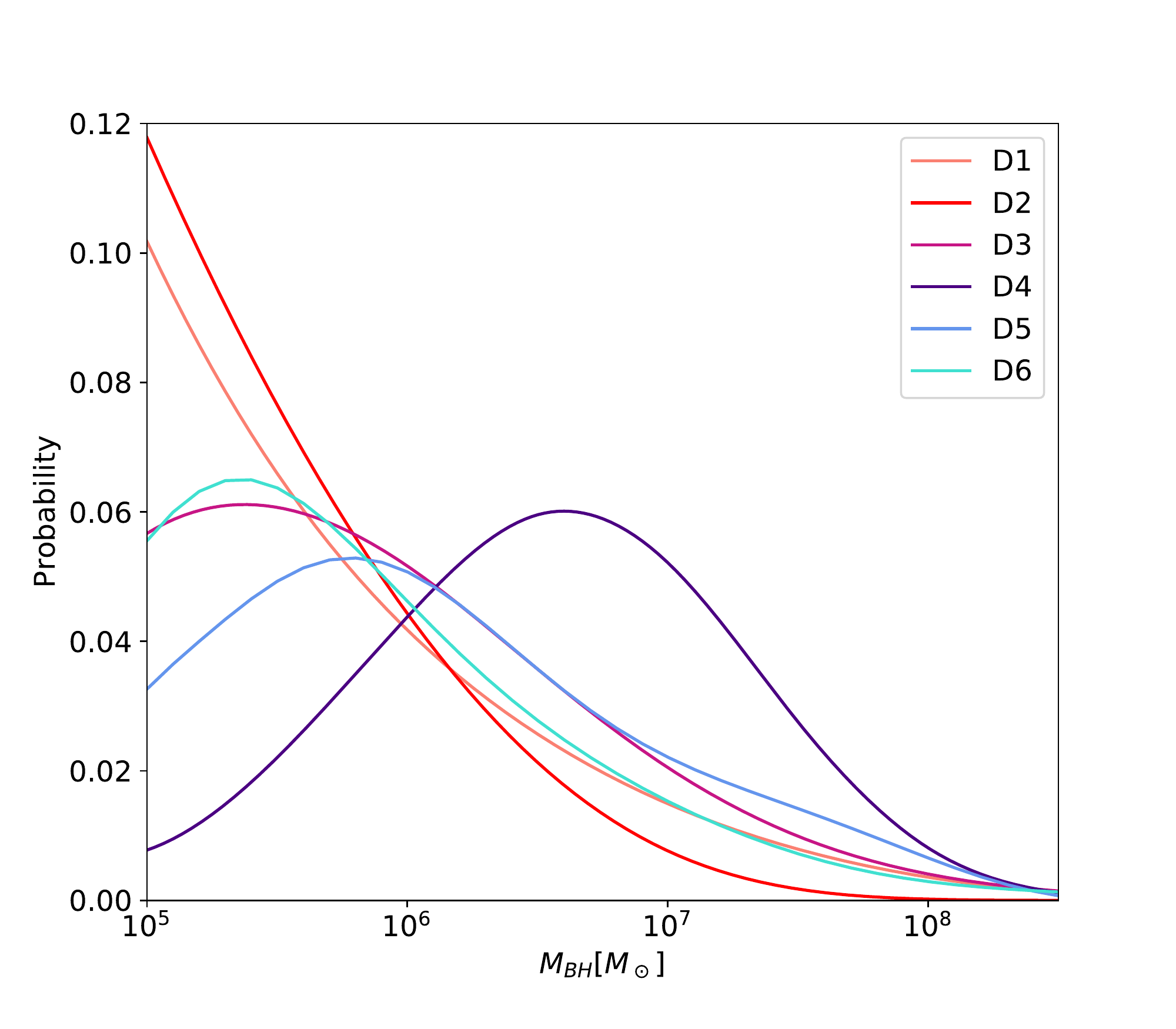} 
\caption{The input SMBH mass probability distributions \texttt{D1}-\texttt{D6}.} 
\label{fig:input_distributions} 
\end{figure}

One of the input parameters for our simulations is the distribution of SMBHs in the centers of galaxies over their masses. Since a SMBH mass influences the brightness and the duration of a TDE, and consequently the chances of its detection, the assumed mass distribution of SMBHs has an impact on the expected TDE detection rates. For our simulations we used six different probability distributions, \texttt{D1}-\texttt{D6}, shown in Figure \ref{fig:input_distributions}. 

Distribution \texttt{D1} was reproduced from \cite{Aversa:2015bya}, where the SMBH mass distribution is described by a Schechter function, given by the equation and parameter values for the BH mass function in their Table 1. \texttt{D2} was reproduced with results from \cite{Hopkins:2006fq}, where the SMBH mass distribution is also described by a Schechter model (equation (24) in \citealt{Hopkins:2006fq}), with its parameter values given in their Table 5. For both distributions we assumed that the parameters describing the distribution do not evolve with redshift. \texttt{D1} and \texttt{D2} only seem to be valid from $10^6 M_{\odot}$ on, however to test the influence of the distribution at $M_{BH} < 10^6 M_\odot$, we have extrapolated them towards the low mass end (down to $10^5 M_{\odot}$).

For \texttt{D3} and \texttt{D4} we used the same function as for describing the distribution \texttt{D1}, however we varied the parameter values of the Schechter function ($\phi$, $X_c$, $\alpha$, $\omega$ in Table 1 of \citealt{Aversa:2015bya}) in such way, that \texttt{D3} peaks around 10$^{5.5} \, M_\odot$ and then slowly falls towards the lower masses, while \texttt{D4} peaks around $10^{6.5} \, M_{\odot}$ and then gradually falls towards the lower black hole masses. The remaining distributions, \texttt{D5} and \texttt{D6}, were calculated using relations for the total stellar mass vs. the black hole mass (\texttt{D5}) and the host galaxy color vs. the total stellar mass (\texttt{D6}). We obtained the total stellar mass and colors of all of the host galaxies from the LSST simulator database and calculated the distributions of black hole masses in the simulator. The relations we used were the following
\begin{equation}
\log \bigg( \frac{M_{BH}}{M_{\odot}}\bigg)  = 1.21 \log 
\bigg( \frac{M_*}{10^{11} M_{\odot}} \bigg) + 8.33
\label{stellarbh}
\end{equation}
for \texttt{D5} \citep{Bosch:2016xik}, where $M_*$ is the total stellar mass in the galaxy, and
\begin{eqnarray}
\log \bigg( \frac{M_*}{M_{\odot}}\bigg) &= 1.097 (g - r) -  
0.4 \bigg(r - 5  \log \frac{d}{10 \, {pc}}\bigg) 
 \nonumber
\\& - 0.19 z + 1.462
\label{colorstellar}
\end{eqnarray}
for \texttt{D6} \citep{Bernardi:2009rf}, where $g$ and $r$ are the magnitudes of the galaxy in \emph{g} and \emph{r} band. We then used equation (\ref{stellarbh}) to calculate the corresponding black hole masses for \texttt{D6}.

\section{Simulations Setup}
\label{simulations}

The LSST's capabilities will enable fast and deep imaging of the whole visible sky on short time scales, which will, among other things, be an important tool for the detection of transient astrophysical phenomena. To understand how different components of the telescope, such as its design, the conditions at the observing site, and the observing strategy will affect the properties of the obtained data, a simulation framework has been designed in order to simulate the whole operation of the telescope \citep[e. g.][]{Ivezic:2008fe, Connolly2014, Peterson2015, Delgado2016}. 

This framework includes a catalog of astronomical objects, \texttt{CatSim}, which contains catalogs of Solar System objects, stars, galaxies, and transients, such as AGNs and micro-lensing events. TDEs have not been included in \texttt{CatSim} yet. The simulator also provides a tool for simulating the operations of the telescope, called \texttt{OpSim}. Together with \texttt{CatSim}, it can be used to simulate observed light curves of various astronomical objects. 

\begin{figure*}
\centering
\includegraphics[width=\textwidth]{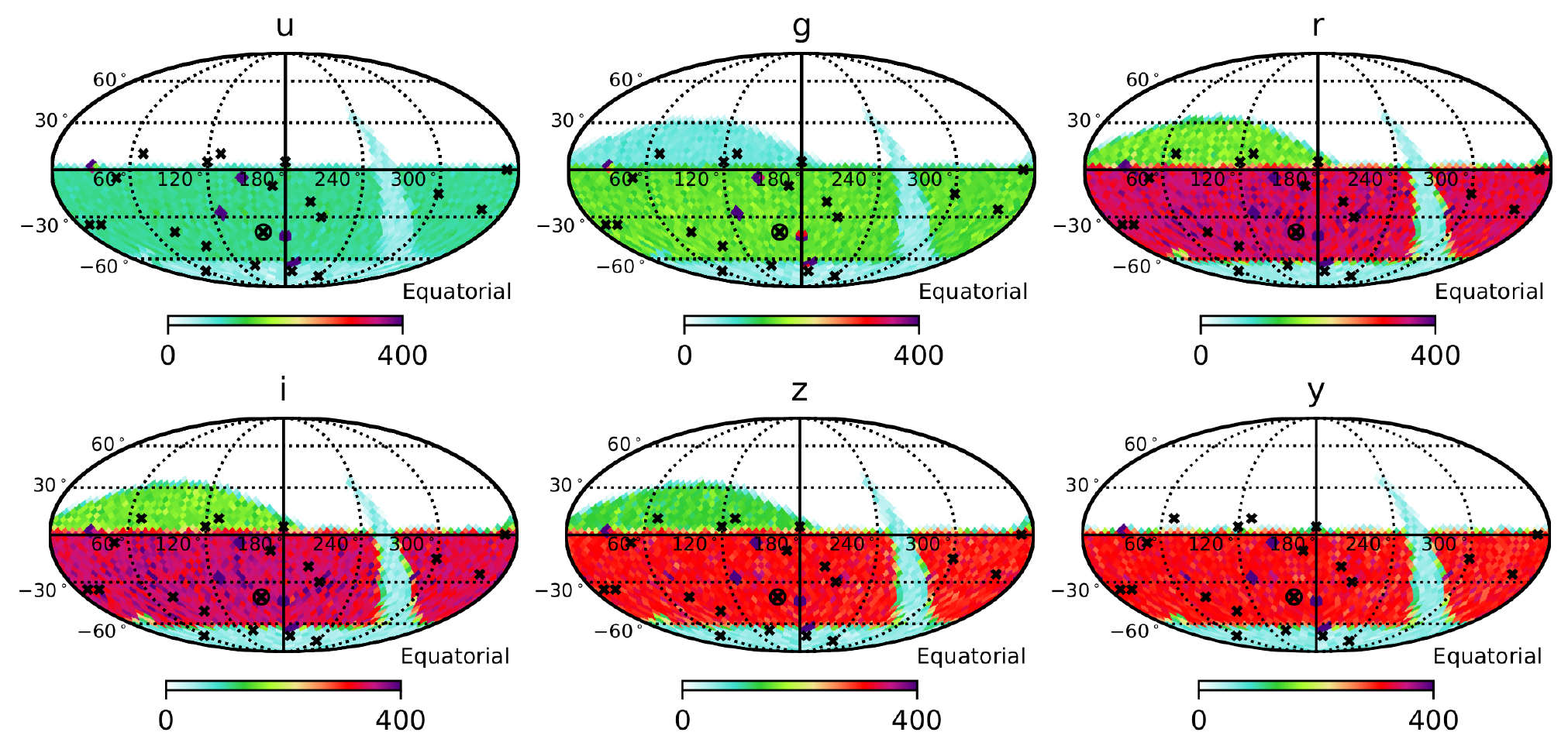} 
\caption{Number of visits (where a visit consists of two 15 second exposures) to a given field on the sky over 10 years of LSST observations in all six bands \emph{u, g, r, i, z} and \emph{y}, according to the observing strategy \texttt{minion\_1016}. Observations in \emph{r, i, z} and \emph{y} band will be more common than those in \emph{u} or \emph{g} band, which is also apparent from the panels corresponding to each of the bands. The distribution of number of visits on the sky is irregular, since the cadence proposed is also irregular. With black crosses we mark the locations of fields on which we simulated TDEs.} 
\label{fig:minion} 
\end{figure*}

The \texttt{OpSim} contains an observation scheduler for the telescope. The observing strategy we used in our simulations was the strategy called \texttt{minion\_1016}\footnote{Recently, the LSST community proposed a number of cadences \citep{IvezicWpCall}, which describe different observing strategies. Before this, the \texttt{minion\_1016} cadence was the baseline strategy of the project, but the exact strategy is yet to be determined in order to satisfy all scientific areas optimally. We tested the new proposed cadences as well, however, for the sole purpose of the number of TDE detections we are estimating here, there are no large discrepancies between \texttt{minion\_1016} and the other cadences (the numbers vary at most by a factor two, see \citealt{Bricman:2018cys}). Therefore, we concluded, that the \texttt{minion\_1016} cadence is representative enough for our purposes here.}, in which each visit of a given field of the sky consists of two 15-second exposures, with the same field being visited again on average every 3 days. The next visit to the same field is scheduled based on the following ranking algorithm \citep{Ivezic2013}: after a visit of a given field, all possible next observations are assigned a score, which depends on their locations, the times of previous observations, and the filters. Therefore, the cadence (i.e. the next visit to the same field) of observations is irregular, and some fields might be visited more frequently than others, and in different filters \citep{Ivezic:2008fe}. For \texttt{minion\_1016}, 7.5\% of the total observing time will be spent observing in \emph{u} band, 10.1\% in \emph{g} band, 22.0\% in \emph{r} band, 22.1\% in \emph{i} band, 20.1\% in \emph{z} band, and 18.2\% in \emph{y} band. The number of visits to a given field of the sky in all six bands over the 10 year survey duration for \texttt{minion\_1016} is shown in Figure \ref{fig:minion}. The average number of visits to a field is 62, 88, 199, 201, 180 and 180 per \emph{u, g, r, i, z} and \emph{y} band, respectively \citep{Marshall:2017wph}. The mean number of visits per field over 10 years is 910.

For our simulations we first generated a catalog of galaxies, which will host a TDE during the 10 years of LSST operation. We then queried the \texttt{CatSim} galaxy database, which covers approximately 20.25 deg$^2$ on the sky and contains around $17$ million galaxies. Since the number of visits changes with respect to the location of the field in the sky, we chose to run our simulations on 20 different fields of size 20.25 deg$^2$. Coordinates of the centers of all 20 fields are marked on Figure \ref{fig:minion} with black crosses.

We randomly chose TDE host galaxies based on the rate of $10^{-5}$ per galaxy per year, and assumed that one galaxy can experience only one TDE in $10$ years of LSST observations. Each host galaxy in the catalog already has defined parameters, such as coordinates, redshift, extinction, etc. However, there is no information on the type of the galaxy. Although there is observational evidence that TDEs prefer E+A post starburts galaxies \citep[e.g. ][]{Arcavi:2014iha, 2016ApJ...818L..21F, Law-Smith:2017zne, Graur:2017vjj}, we did not differentiate between the different types of galaxies in our simulations. We assumed that all of the galaxies are equally likely to host a TDE and used the TDE rate ($10^{-5}$ per galaxy per year) inferred from observations, which is averaged over all galaxy types. This implies an assumption that E+A galaxies are distributed isotropically across the sky (as other types of galaxies).

There is also no information on the mass of the host galaxy's central SMBH in the \texttt{CatSim} catalog, therefore we assigned it randomly from an assumed SMBH mass distribution (we consider 6 different distributions presented in Section \ref{smbhdistribs}). Note that the black hole masses were randomly chosen from an interval between $10^5 \, M_\odot$ and $10^8 \, M_\odot$, since black holes with masses larger than $10^8 \, M_\odot$ will swallow a Solar type star before it gets disrupted (the tidal radius would be within the Schwarzschild radius of the black hole).

For each SMBH mass distribution discussed in Section \ref{smbhdistribs}, we created 20 TDE host galaxy catalogs, one for each simulated patch on the sky. Each of the catalogs contained around 1700 host galaxies, including active galaxies (approximately $1$\% were AGNs), which we have eliminated from further investigation, since the characteristics of TDEs happening inside AGNs are not known. 

We assigned each host galaxy a TDE with a starting time drawn randomly from the duration of the survey. In all cases we assumed that the disrupted star is Solar-like ($M=M_\odot$, $R=R_\odot$, polytropic model that blends between $\gamma=4/3$ and $\gamma=5/3$). We randomly assigned each disruption a $\beta$ value, which we let vary from $\beta_{min} = 0.6$ to $\beta_{max} = 11.8 M_{BH,6}^{-2/3}$ ($M_{BH,6}$ is the mass of the black hole in units of $10^6 M_\odot$). The value of $\beta_{max}$ corresponds to a pericenter at 2 Schwarzchild radii. We assume that a star which travels any further into the black hole's gravitational potential does not produced a bright flare before it is swallowed by the black hole. The values of $\beta$ between $0.6$ and $1.8$ correspond to a partial disruption, and values larger than $1.85$ correspond to a full disruption of the star \citep{Guillochon:2012uc}. Assuming that the probability for an encounter with a pericenter distance between $r_p$ and $r_p + dr_p$ is proportional to the area $2\pi r_p dr_p$, we distributed $\beta$ according to the following function
\begin{equation}
p(\beta) = \frac{1}{2 \, \beta^3} \bigg(\frac{1}{\beta_{min}^2} -
 \frac{1}{\beta_{max}^2}\bigg)^{-1},
\end{equation}
 making disruptions with smaller $\beta$ values more probable than those with larger penetration factors. Since \cite{Guillochon:2012uc} have noticed that $\beta$ values larger than $4.0$ do not produce any substantial change in the behavior of the fallback rate, and consequently in the behavior of the light curve and the SEDs, we assumed that the SED stays the same as in the case of $\beta = 4.0$, if $\beta$ value was larger than $4.0$.

For every TDE in the catalog the flux was calculated using \texttt{MOSFiT} and applying the cosmological redshift of the host galaxy to the SED\footnote{We assumed a flat Universe with cosmological parameters $\Omega_0 = 0.25$, $\Omega_\Lambda = 0.75$ and $H_0=73$ km/s/Mpc.}. During the simulations, the host galaxy and the Milky Way dust extinctions were applied to each event according to the model in \cite{ODonnell1994}. 

Simulations of light curves were done in all six LSST bands, for galaxies with redshifts $z<3.0$, since events at larger redshifts are expected to be too dim to be observed. Using host galaxy R.A. and Dec, we queried the \texttt{minion\_1016} database, which contains a simulated observing cadence for the LSST, based on the algorithm described above. At each time a certain TDE in the sky was observed, its magnitude in a given band was calculated along with an error-bar. 

\begin{figure}
\centering
\includegraphics[width=.92\linewidth]{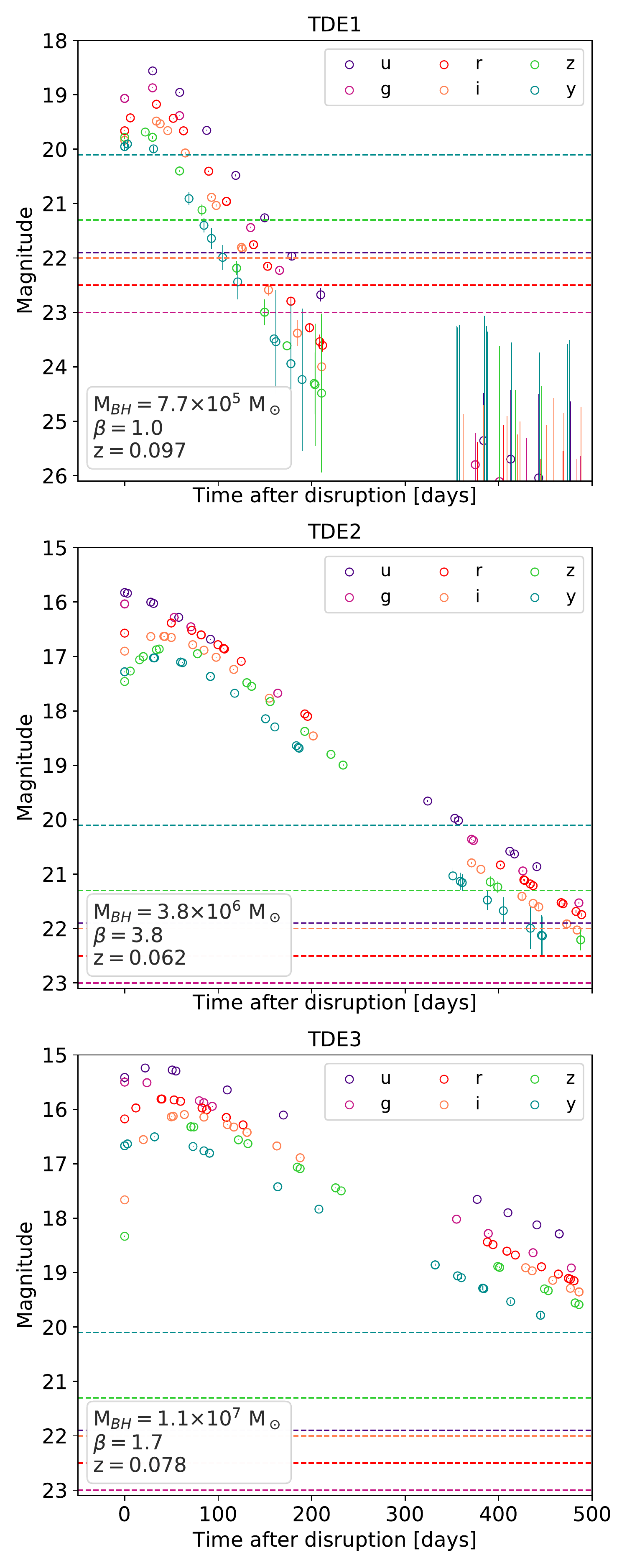} 
\caption{Simulated observed light curves of three different $1M_\odot$ disruptions in all six LSST bands (\emph{u, g, r, i, z} and \emph{y}). TDE1 is a disruption with redshift at $z=0.097$, $M_{BH} = 7.7 \times 10^5 M_{\odot}$ and $\beta = 1.0$, TDE2 is at $z = 0.062$, $M_{BH} = 3.8 \times 10^6 M_{\odot}$ and $\beta = 3.8$, and TDE3 is at $z=0.078$,  $M_{BH} = 1.1 \times 10^7 M_{\odot}$ and $\beta = 1.7$. Error-bars (vertical lines) are also plotted together with cut-off magnitudes (as defined in \ref{results_detection}) for each filter (horizontal dashed lines).} 
\label{fig:lightcurves} 
\end{figure}

Examples of obtained light curves of three simulated events are shown in Figure \ref{fig:lightcurves}. The events have different parameters and are at different redshifts. TDE1 is at $z=0.097$, where the disrupting black hole has a mass of $M_{BH} = 7.7 \times 10^5 M_{\odot}$, and the penetration factor is $\beta = 1.0$. TDE2 is at $z=0.062$, with $M_{BH} = 3.8 \times 10^6 M_{\odot}$ and $\beta =3.8$, while TDE3 is at $z=0.078$ with $M_{BH} = 1.1 \times 10^7 M_{\odot}$ and $\beta = 1.7$.

Note that our simulations do not contain any deep drilling fields, which the LSST is expected to spend $10$\% of the observing time on. All of the fields chosen for our simulations are within the wide-fast-deep area of the observing strategy.

\section{Results and Discussion}
\label{results}

\subsection{Detection definition}
\label{results_detection}

\begin{figure*}
\centering
\includegraphics[width=.9\textwidth]{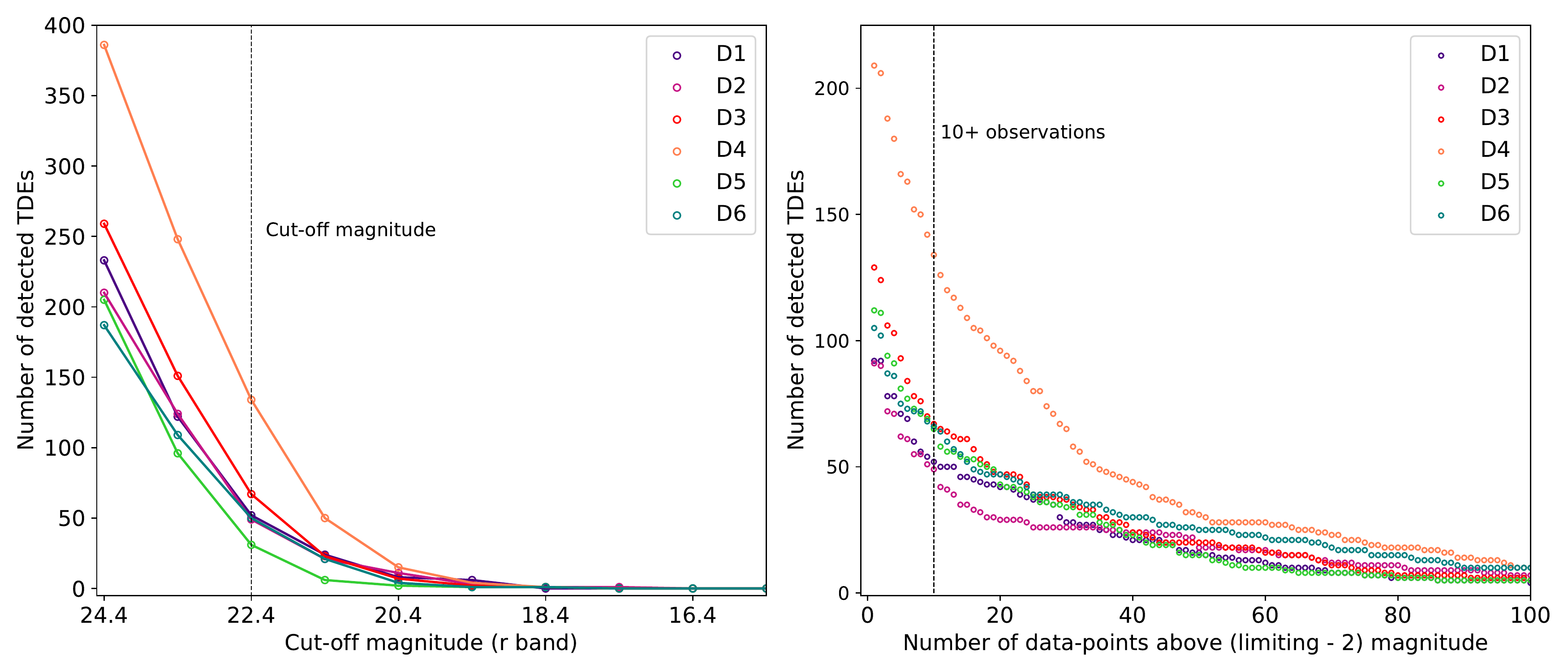} 
\caption{\emph{Left}: the number of TDEs seen at least once above a certain cut-off \emph{r}-band magnitude over 10 years of LSST observations on 20.25 deg$^2$ of the sky as a function of the cut-off magnitude in \emph{r} band. The number of detected TDEs decreases as we go to brighter limits. We chose (limiting $-$ 2) magnitude to eliminate events close to the limiting magnitudes of each band. \emph{Right}: the number of detected TDEs over 10 years of LSST observations on a small patch of 20.25 deg$^2$ of the sky as a function of the number of data-points above the chosen cut-off magnitude, (limiting magnitude $-$ 2), in all LSST bands together. For a representative number, we assumed 10 observations above the cut-off magnitude as sufficient to classify the event.} 
\label{fig:detection_curves} 
\end{figure*}

To estimate the number of TDE detections, we first need to define what counts as a detection. Too few data points or data points very close to the limiting magnitude (with large error-bars) do not assure a positive identification of the source as a TDE. In Figure \ref{fig:detection_curves} (left) we plot the number of TDEs seen at least once above a certain magnitude, which we call the cut-off magnitude, over 10 years of LSST operations, simulated on one patch of 20.25 deg$^2$ on the sky (in Figure \ref{fig:minion} marked with $\otimes$). As expected, the fainter the cut-off magnitude, the more events will be observed, however data points close to the limiting magnitude will have large error-bars. Setting a brighter cut-off magnitude reduces the number of detected events, but on the other hand it means that data points will carry smaller error-bars and result in better quality light curves. We decided to set the cut-off magnitude to the (limiting $-$ 2) magnitude of the band, so that the cut-off magnitudes in the remainder of this paper are $u_c = 21.5$, $g_c = 22.8$, $r_c = 22.4$, $i_c = 21.9$, $z_c = 21.3$ and $y_c = 20.1$. 

For the positive identification of a TDE, based only on the LSST data, it will be important how many good quality data points the light curve will contain, i.e. how many times a TDE is detected above the cut-off magnitude. Figure \ref{fig:detection_curves} (right) shows the number of TDEs on a patch $\otimes$ (Figure \ref{fig:minion}) seen above the cut-off magnitude (as defined in the previous paragraph) over 10 years of LSST operations at least a certain number of times, given on the x-axis. As expected, the fewer points we choose as sufficient for a positive identification, the higher the number of events. The plot does not show any clear trend which would suggest what boundary would be the best choice. We arbitrarily chose 10 data points across all LSST bands as a minimum number sufficient for a reliable classification of a TDE. This may not be sufficient to distinguish a TDE from other types of transients, especially supernovae, however observations in different bands on a short-time scale (e.g. day or two) could provide us with color information, which could be helpful in that. We note that this number might be different once a well performing classification tool for identifying TDEs out of a large number of transients is produced.

\subsection{Number of TDEs detected}
\label{results_detectedTDEs}

\begin{figure}
\centering
\includegraphics[width=\linewidth]{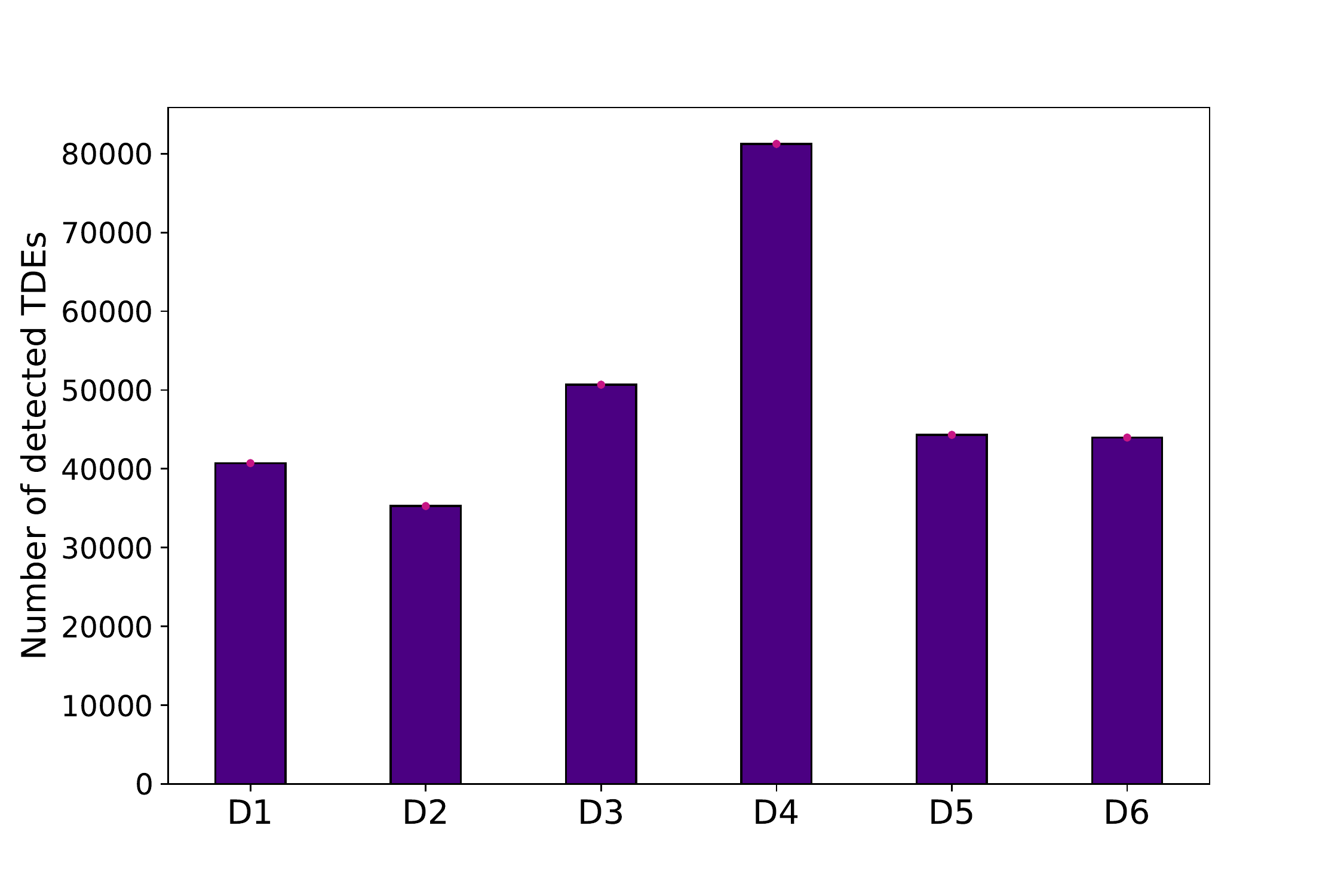}
\caption{The number of detected TDEs for each of the SMBH mass distributions \texttt{D1-D6}. The number of expected detections is between 35 000 $\pm$ 260 and 80 000 $\pm$ 400 over 10 years of observations. This corresponds to average values between 10 and 22 TDEs per night.}
\label{fig:detectedTDEs}
\end{figure}

To calculate the number of TDEs observed over the whole LSST visible sky, we first divided it into three areas, which have significantly different number of visits, as clearly evident in Figure \ref{fig:minion}. We put 4 of the simulated fields in area I (Dec $ > 0^\circ$, size $\sim 3300$ deg$^2$), 4 fields in area II (Dec $ < -60^\circ$, size $\sim 1700$ deg$^2$), and 12 fields in area III ($-60^\circ <$ Dec $ < 0^\circ$ excluding the galactic plane, size $\sim 13\, 000$ deg$^2$). On the total of 20 patches in these three areas we performed simulations and calculated the mean number of detections for all six SMBH mass distributions described in Section \ref{smbhdistribs}. To obtain the total number of TDEs over the whole LSST visible sky, we weighted the mean numbers obtained for these patches with their area size and summed all of the contributions. Figure \ref{fig:detectedTDEs} shows the mean number of detected TDEs for each of the SMBH mass distributions.

The uncertainties were estimated by first calculating the standard deviation of the detected TDEs for 12 small patches in the sky area III. The standard deviations of the number of detections were negligible in the other two areas I and II compared to area III. From our results for area III we find that the standard deviations are $\sim$1.4 $\sqrt{N}$, where $N$ is the number of detected TDEs. We used this approximate relation to estimate the uncertainties of the number of detected TDEs in the whole sky. The uncertainties are shown in Figure \ref{fig:detectedTDEs} with pink and are very small compared to the total number of TDEs.

The number of detected TDEs largely depends on the underlying SMBH mass distribution, as well as on the choice for the cut-off magnitude. In our case, the number of detected TDEs lies in an interval roughly between 35 000 $\pm$ 260 and 80 000 $\pm$ 400 events over 10 years of LSST observations. This corresponds to roughly 10 to 22 TDEs on average per night.

We would like to stress here that the number of detected TDEs strongly depends on the cut-off magnitude and the minimum number of data points above this magnitude required for a positive TDE detection as discussed in \ref{results_detection}. Choosing the cut-off magnitude to be (limiting $-$ 3) magnitudes instead of (limiting $-$ 2) magnitudes, gives a number of detected TDEs between 27 000 $\pm$ 230 and 65 000 $\pm$ 350 over 10 years, corresponding to 7 to 18 TDEs on average per night. Similarly for the number of data points required for a detection: requiring 20 data points above the cut-off magnitude instead of 10, lowers the number of detected TDEs over 10 years to be between 23 000 $\pm$ 140 and 58 000 $\pm$ 360, or an average of 6 to 16 TDEs per night.

The number is the highest for the SMBH mass distribution \texttt{D4}, since the peak of the distribution is at a higher black hole mass ($\sim 10^{6.5} \, M_\odot$) and it falls rapidly towards the low-mass end. In general, TDEs at the low mass end are less luminous and decline more rapidly with time, therefore the probability of missing them with an irregular cadence is high. In the case of \texttt{D4} mass distribution the number of TDEs at the low mass end is small and non-detection of these events does not have a large effect on the overall number of observed events. The brighter events caused by more massive black holes are in this case more frequent and more efficiently detected, therefore the total number of detected TDEs is larger. 

The distribution with the smallest number of detected TDEs is \texttt{D2}, however the number of detected TDEs in this case is similar to the number of detected TDEs in all other distributions (except \texttt{D4}). The reason for the low number of detected TDEs in \texttt{D2} is that, as seen in Figure \ref{fig:input_distributions}, \texttt{D2} has the largest number of black holes at the low mass end compared to other distributions. Therefore, the majority of the events will be caused by less massive black holes, consequently TDEs will be less luminous, fade quicker, and the probability of not detecting them will be greater.

Our results are in rough agreement with \cite{vanVelzen:2010jp} and \cite{Abell:2009aa}. \cite{vanVelzen:2010jp} have estimated, based on the previous observations, that the LSST should discover around 40 000 new TDEs over 10 years, while in \cite{Abell:2009aa} this number was estimated to be 60 000, based on the universal TDE rates from \cite{Rau:2009yx} and the references therein. 

It is worth mentioning here that we assume all of the detected TDEs will be recognized and classified as TDEs, which might not be true in all cases. Due to a large number of transients the LSST is expected to discover every night ($\sim$ 10 000), distinguishing TDEs from other transients will be a difficult task and probably not always straight-forward. Some events might be mis-classified as supernovae or as active galactic nuclei, which could largely affect the total number of detected TDEs. There are some observational features, such as the distance from the galactic center, the color (TDEs tend to be very blue in color, with $g-r \approx -1.0$), the light curve shape, and the color and temperature evolution, which should be helpful in distinguishing TDEs from other types of transients, however observations at other wavelengths or spectra will probably be needed for reliable classification.

The availability of follow-up observations will depend on TDEs redshifts. Taking redshifts $z<$ 0.2 (the limit set by the current observed sample of TDEs, which mostly have redshifts $<$ 0.2) we find from our simulations that approximately 10\% to 20\% of the detected TDEs lie within this range and would therefore be good candidates for follow-up observations.

Making constraints in addition to $z<$ 0.2, further reduces the number of detected TDEs. Requiring that a TDE should have at least 2 observations in any band before the peak (to be able to accurately determine the peak time and peak magnitude), and at least 5 observations in any band within 30 days after the peak, results in only 5\% to 8\% of the whole detected sample (depending on the SMBH mass distribution), giving a number of detected TDEs between 2500 and 3000 over 10 years of observations.

In order to be able to fully exploit the LSST data, a method for the identification of TDEs from photometric data alone is needed. This problem, however, is outside the scope of this work. In the following discussion, we will assume that all the detected TDEs will be classified.

\subsection{Probing the SMBH mass distribution}
\label{results_smbhs}

Once the LSST starts observing TDEs on a daily basis, the masses of black holes responsible for causing TDEs could be determined by fitting the observed light curves with a TDE light curve model. Provided that the events have good sampling around the peak or observations in various bands, as shown in \cite{Mockler:2018xne}, the black hole masses can be well constrained. In addition, even if there are no data points near the peak, but there is good sampling of the decline, the black hole masses can still be determined. \texttt{MOSFiT} produces a number of light curves that can fit an event, and consequently a number of different estimations on the SMBH mass. The statistical error-bars on the SMBH masses determined in \cite{Mockler:2018xne} are small, within 0.1 dex and only in some cases as large as 0.3 dex. The SMBH masses determined in their work are in agreement with SMBH masses determined with the $M_{BH}-\sigma$ relation. In addition, their results show no clear correlation between any other parameters (such as for example $\beta$ with the black hole mass).

\begin{figure*}
\centering
\includegraphics[width=.95\textwidth]{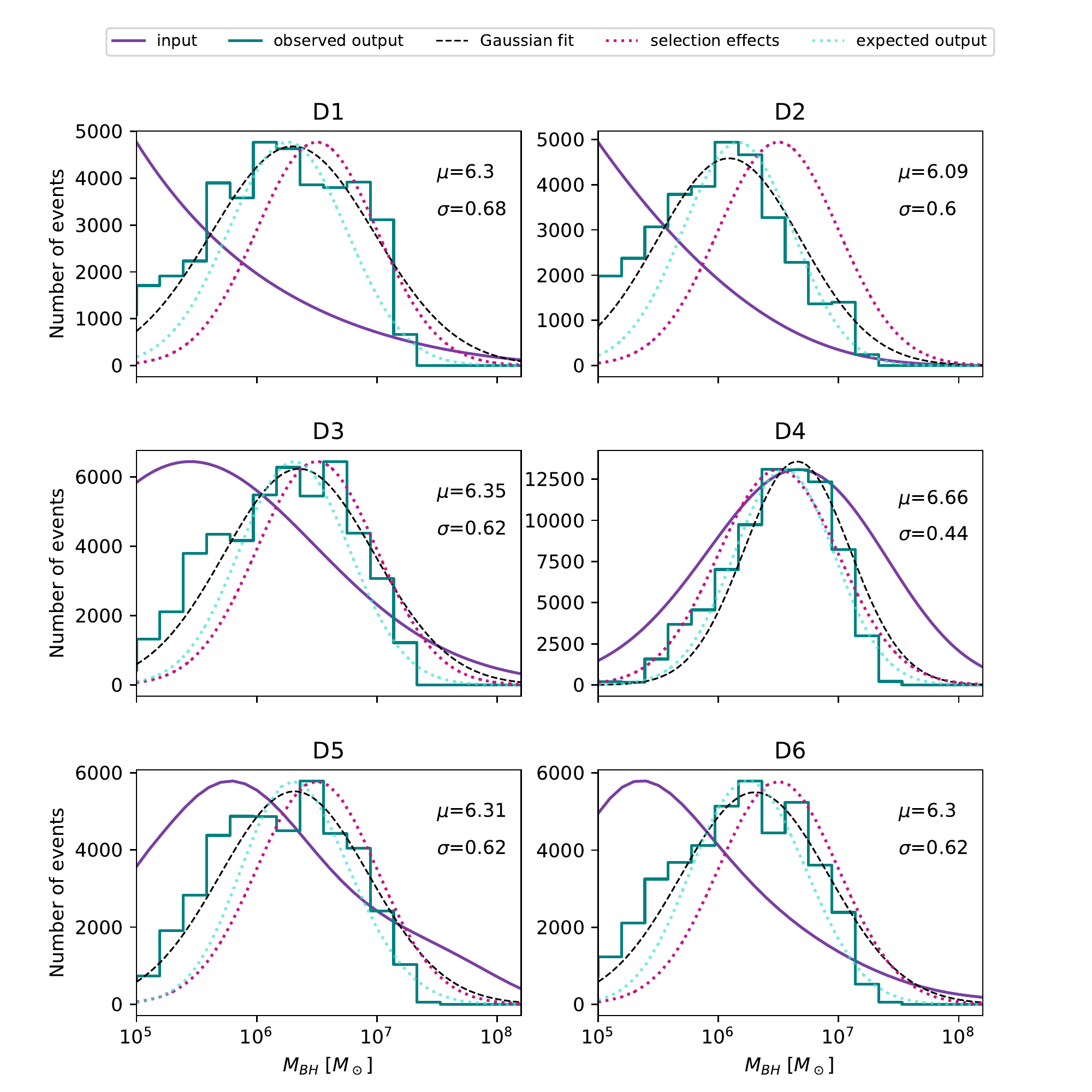} 
\caption{The input theoretical SMBH mass distributions \texttt{D1-D6} (purple lines), the SMBH mass distribution of detected TDEs (green histograms), the Gaussian fit to detected TDEs (black dashed line) and the selection-effects function (pink dotted line). The observed samples consist of all detected TDEs on 20 simulated fields, scaled to the whole observable sky as discussed in \ref{results_detectedTDEs}.}
\label{fig:final_distributions} 
\end{figure*}

To address the possibility of probing SMBH mass distributions with TDEs observed by the LSST, the distributions of detected TDEs over black hole mass together with the initial input distribution for all six initial distributions are shown in Figure \ref{fig:final_distributions}.

From Figure \ref{fig:final_distributions} it is clear that none of the simulated TDE distributions follow the initial distribution of SMBH masses. At the high mass end we notice that the distributions of detected TDEs over black hole mass seem to follow the initial distributions quite well, however the distributions fall quickly towards $10^8$ $M_\odot$, since a Solar type star enters a heavier black hole before it can be disrupted, and no flare is observed.

At the low mass end, however, none of the detected TDE distributions over SMBH masses follow the initial distributions. This is due to the fact that TDEs involving a less massive black hole produce less luminous events, which fade faster with time (see Figure \ref{fig:bh_comparison}), making them harder to detect. It is possible that we are missing those dim and short TDEs due to the cadence we used in our simulations.

To describe the effect of observational bias, i.e. the faintness and short duration of events at the low-mass end, we created a ``selection-effect'' function (pink dotted lines in Figure \ref{fig:final_distributions}). We simply assumed that it is a Gaussian function, the same for all SMBH mass distributions, with a mean value of $10^{6.5} \, M_\odot$ and a standard deviation of 0.5 on a logarithmic scale. We then multiplied  this function with the initial input distribution (purple line in Figure \ref{fig:final_distributions}) and noticed that the final product fits the observed TDE distributions well (green histograms in Figure \ref{fig:final_distributions}). This product is shown in Figure \ref{fig:final_distributions} with dotted light blue lines.

To check if it would be possible to distinguish between different initial distributions, we fitted a Gaussian function to the detected TDE distributions (black dashed lines in Figure \ref{fig:final_distributions}). We find that in all cases the mean and the standard deviations (also noted in each panel of Figure \ref{fig:final_distributions}) have very similar values and cannot be used to reach reliable conclusions regarding the initial distributions.

From the results obtained in our simulations it seems that it will not be straightforward to deduce the mass distribution of SMBHs in spite of a large number of TDEs detected by the LSST. This is due to the selection effects which, provided that the theoretical models are giving correct predictions about the duration and luminosity of TDEs, are biased against low-mass black hole TDEs.

In principle, the total number of detected TDEs could tell us something about the shape of the SMBH mass distribution. As mentioned in Section \ref{results_detectedTDEs}, different SMBH mass distributions give different total numbers of TDEs detected. However, the number of detected TDEs strongly depends also on the rate of TDEs, which is not yet firmly known. Therefore, until the rate of TDEs is more precisely known, it will not be possible to lift the degeneracy between the rate and the SMBH mass distribution, and use the total number of detected TDEs as a strong indicator for the shape of the SMBH mass distribution.

We would like to note that our results are obtained with the \texttt{minion\_1016} cadence. We tested whether simulations over the whole observable sky might improve the statistics at the low mass end of the SMBH distribution. We find that running simulations on a larger number of fields (e.g. 20 instead of 10) does not change the shape of the resulting mass distribution of detected TDEs significantly and it only slightly affects the $\mu$ and $\sigma$ of the Gaussian fit. We also tested the effects of different requirements in our definition of a TDE detection, i.e. the cut-off magnitude and the number of data points above the cut-off magnitude. We find that changing these two parameters does not affect the shape of the mass distribution of detected TDEs significantly. The low-mass end sampling might be better using a different cadence (e.g. one of the cadences mentioned in \citealt{Bricman:2018cys}), which has a more frequent temporal sampling.

\section{Conclusions}
\label{conclusions}

Based on results in Figure \ref{fig:detectedTDEs} we estimate the LSST will discover on average between 35 000 and 80 000 TDEs over 10 years of observations, depending on the SMBH mass distribution. This corresponds to approximately 10 to 22 TDEs on average per night. We may therefore expect that the LSST will significantly enlarge the sample of observed TDEs, improve the statistics concerning their properties, and our understanding of these transients.

These numbers are based on the assumption that 10 data-points in any LSST band will be sufficient for a TDE classification. Requiring 20 data-points above the cut-off magnitude in any band reduces the number of detections to be between 23 000 $\pm$ 140 and 58 000 $\pm$ 360 over 10 years, or an average of 6 to 16 TDEs per night. An additional constraint of a TDE distance $z<$ 0.2 (to allow for follow-up photometric and spectroscopic observations with mid-size telescopes) reduces the number to only around 10\% of the detections inferred from Figure \ref{fig:detectedTDEs}. Therefore, to reliably classify a majority of the TDEs detected by the LSST, a photometric classifier is of crucial importance.

The distributions of detected TDEs over the black hole masses involved in the process are not as informative about the underlying SMBH mass distribution as one might hope. Based on the results in Figure \ref{fig:final_distributions}, there is no clear parameter with which we could distinguish among different initial distributions. We find that this is a consequence of the short duration and faintness of TDEs caused by low-mass SMBHs, due to which the majority of such TDEs might be missed by observations.

We expect that a cadence with a more regular or more dense sampling might give a higher number of detected events at the low mass end of the SMBH mass distribution, providing additional information on the mass distribution of SMBHs.

In any case, a substantially larger sample of TDEs detected by the LSST and supplemented by follow-up observations with other facilities, promises to provide many new insights into the properties of TDEs including their stellar and black hole properties (mass, spin), the ambient environment of quiescent SMBHs, their host galaxies properties, and last but not least, more strict limits on the true rate of TDEs.

\acknowledgments

We thank Brenna Mockler for very helpful discussions concerning \texttt{MOSFiT}. We thank Scott Daniel for helping us set up the LSST simulation framework and for his helpful insights about the light curve generation process. We acknowledge the financial support from the Slovenian Research Agency (research core funding P1-0031, infrastructure program I0-0033, project grant No. J1-8136, and KB's Young Researcher grant) and networking support by the COST Action GWverse CA16104.

{\software{\texttt{MOSFiT} \citep{Guillochon:2017bmg},  
          LSST Catalogs (\texttt{CatSim}) \citep{Connolly2014},  
          Operations Simulator (\texttt{OpSim}) \citep{Delgado2016},
          Matplotlib \citep{Hunter:2007ouj}, 
          Scipy \citep{JonesScipy}}}

\bibliographystyle{aasjournal}
\bibliography{references}

\end{document}